\documentclass[fleqn,10pt]{wlscirep}
\usepackage[utf8]{inputenc}
\usepackage[T1]{fontenc}
\usepackage{physics}
\usepackage{bm}
\usepackage{qcircuit}
\usepackage{comment}
\usepackage{here}
\usepackage{indentfirst}

\newcommand{\arrep}[1]{\ar@{-}@/^/[#1]|-{\mbox{ $L$ times }}}

\title{Application of time-series quantum generative model to financial data}

\author[1]{Shun Okumura}
\author[1, 2, 3,*]{Masayuki Ohzeki}
\author[4]{Masaya Abe}
\affil[1]{Graduate School of Information Sciences, Tohoku University, Miyagi 980-8564, Japan}
\affil[2]{Department of Physics, Tokyo Institute of Technology, Tokyo, 152-8551, Japan}
\affil[3]{Sigma-i Co., Ltd., Tokyo, 108-0075, Japan}
\affil[4]{Nomura Asset Management Co., Ltd., Tokyo, 135-0061, Japan}
\affil[*]{mohzeki@tohoku.ac.jp}

\begin{abstract}
Despite proposing a quantum generative model for time series that successfully learns correlated series with multiple Brownian motions, the model has not been adapted and evaluated for financial problems.
In this study, a time-series generative model was applied as a quantum generative model to actual financial data.
Future data for two correlated time series were generated and compared with classical methods such as long short-term memory and vector autoregression. 
Furthermore, numerical experiments were performed to complete missing values.
Based on the results, we evaluated the practical applications of the time-series quantum generation model.
It was observed that fewer parameter values were required compared with the classical method.
In addition, the quantum time-series generation model was feasible for both stationary and nonstationary data.
These results suggest that several parameters can be applied to various types of time-series data.

\end{abstract}
\begin{document}

\flushbottom
\maketitle
%
%
\thispagestyle{empty}

\section*{Introduction}

A sequence of data $ \{x_t\}^{T}_{t=1}$ observed between times $t=1$ and $t=T$ is referred to as time-series data.
These data are available in various fields, including finance.
An important topic in this area is analyzing time-series data, such as stock prices, for investment and risk management purposes.
Many time-series datasets behave similarly to random walks \cite{stock}.
Consequently, predicting future data from time-series data is difficult.
However, with recent developments in computer science, many methods have been proposed to predict future values by successfully capturing the correlations between each time series.
The autoregressive (AR) model is the most fundamental model. 
A regression analysis is performed using data from a specified time interval. Next, predictions are made based on the regression results. By iterating this process, predictions can be made up to a certain number of steps.
Recurrent neural networks (RNNs) have also been proposed \cite{RNN}. RNN can improve prediction accuracy via the nonlinear transformation of models, such as the AR model.
However, RNNs are incapable of long-term memory because of gradient loss and other problems \cite{RNNgrad}. Therefore, long time-series data cannot be used for training.
Many models have been proposed that solve this problem; the long short-term memory (LSTM) is a typical model \cite{LSTM}.
In addition, significant developments have occurred in methods aimed at learning the probability distribution governing the data, called generative models.
In this context, generative models have successfully forecast time-series data \cite{GM}.

Recently, research has been conducted on quantum computers based on quantum mechanics. Quantum computers often perform calculations faster than classical computers. Prime factorization and inverse matrix calculations are typical examples\cite{Shor, HHL}. 
Therefore, quantum computers can contribute to the development of machine learning in the future. 
However, these computers are still under development, and few computational resources are available. 
Consequently, large-scale calculations are hindered owing to noise during computation. 
Therefore, active research has been conducted on small-scale quantum computers called intermediate-scale quantum devices (NISQ)\cite{NISQ}.
We expect that NISQ can efficiently generate probability distributions that are difficult to generate using classical computers\cite{NISQprob}.
However, it is uncertain whether real-world valuable algorithms such as prime factorization can be accelerated using NISQ.

NISQ is often used in connection with classical computers because of its limited computational resources.
Quantum machine learning using this hybrid approach has been previously proposed\cite{PSM, QNN}.
Several studies have reported that quantum machine learning can achieve the same prediction accuracy with fewer parameters than that required for machine learning on a classical computer\cite{QGEN}. 
Approaches that forecast time-series data using quantum machine learning also exist.
The quantum generative model is promising owing to its ability to generate probability distributions, which is difficult in the case of NISQ. 
Quantum generative models have been suggested to reduce the number of parameters compared to classical methods by exploiting entanglement, a property unique to quantum mechanics\cite{QGEN}.
In addition, \ cite{copula} proposed a quantum generative model of copulas used in finance.
A quantum generative model for time series has also been proposed that successfully learns correlated series with multiple Brownian motions\cite{tQGEN}.
However, this time-series quantum generative model has not been adapted and evaluated for financial problems.
In this study, we adapted this time-series quantum generation model to correlate time series in finance.
We discuss the performance and practicality of its use in predicting the future and complementing missing values.

This paper is divided into four parts. First, a background is provided in the introduction.
Subsequently, the necessary knowledge, including the time-series quantum generative model, is introduced.
Thereafter, numerical experiments are presented, wherein the time-series quantum generative model is used to predict the future and complement missing values, and the results are discussed.
Finally, we summarise the study and discuss future developments.

\section*{Preliminaries}

\subsection*{Quantum generative model}

\subsubsection*{Quantum Circuit Born Machine}

Quantum circuit Born machines (QCBM) are fundamental to quantum generative models \cite{QCBM}.
This model follows Born's rules of quantum mechanics.
The probability of observing a quantum state is learned, and sampling is generated accordingly.
The quantum state $\ket{\psi}$ is defined based on Born's rule as follows:
\begin{align}
\ket{\psi} = c_{00\dots 0}\ket{00 \dots 0} + c_{00\dots 1}\ket{00 \dots 1} + \dots + c_{11\dots 1}\ket{11 \dots 1}
\end{align}
where $c_{\bm{b}} \in \mathbb{C}$ denotes the complex probability amplitude.
Therefore, $p_{\bm{b}}:=\abs{c_{\bm{b}}}^2$ is the probability of observing the quantum state $\ket{\bm{b}}$ and satisfies $\sum_{\bm{b}}p_{\bm{b}} =1$.
In addition, $\{\ket{00 \dots 0}, \ket{00 \dots 1}, \dots ,\ket{11 \dots 1}\}$ represents the computational basis.
The probability $p_{\bm{b}}$ of obtaining a quantum state $\ket{\bm{b}}$ can be explicitly described as 
\begin{align}
p_{\bm{b}} = \Tr[\ketbra{\psi}{\psi} P_{\bm{b}}]
\end{align}
where $P_{\bm{b}} = \ketbra{\bm{b}}{\bm{b}}$ denotes the projection operator.
Based on this probability, the classical bit sequence $\bm{b}$ is obtained from the measurement.
We parameterized the quantum state $\ket{\psi{(\bm{\theta})}}$ by using a quantum computer to learn the probabilities.
This can be described using the parameterised unitary operator $U(\bm{\theta})$ and the initial state $\ket{\psi_0}$ as follows:
\begin{align}
\ket{\psi{(\bm{\theta})}} = U(\bm{\theta})\ket{\psi_0}.
\end{align}
If the parameterized unitary operator $U(\bm{\theta})$ satisfies certain conditions, optimization may become difficult \cite{BP }.
Therefore, caution should be exercised when designing $U(\bm{\theta})$.
Parameterized probabilities $p_{\bm{b}}{(\bm{\theta})}$ can be estimated by measuring the parameterized quantum circuit and sampling bit strings.
This probability is used for learning by calculating and minimizing the KL divergence with the target probability $q_{\bm{b}}$.
The derivative for optimizing the parameters can be easily computed using the parameter shift rule \cite{PSM}.

\subsubsection*{Time-series quantum generative model}
Several models have been proposed for time-series analysis in quantum machine learning. 
The time-series generation model proposed in \cite{tQGEN} was used in this study.
Open quantum systems inspired this model, and stochastic transition matrices were learned.
Its structure facilitates the encoding of the time series in a quantum circuit, thus providing excellent explanatory properties for the model.
We considered a discrete state $s \in S=\{1, 2, \dots, m\}$ and assumed transitions to elements within $S$ with a certain probability.
The stochastic transition matrix revealed the probability distribution of a given initial state $s_0 \in S$ after $k$ steps.
Using the state probability distribution corresponding to the initial state $s_0$ and the stochastic transition matrix $T$, the state probability distribution after $k$ steps is expressed as follows:
\begin{align}
\mqty(p^{(k)}_1 \\ p^{(k)}_2 \\ \vdots \\p^{(k)}_m) = T^k \mqty(p^{(0)}_1 \\ p^{(0)}_2 \\ \vdots \\p^{(0)}_m)
\end{align}
where $p^{(l)}_i$ is the probability of obtaining state $i$ after $l$ steps.
This relationship was applied to QCBM.
Using the parameterized unitary operator, the quantum state obtained after $k$ steps from the initial quantum state $\ket{\psi_0}$ is
\begin{align}
\ket{\psi{(\bm{\theta})}}  = U^k(\bm{\theta})\ket{\psi_0}.
\end{align}
where $U^k(\bm{\theta})$ corresponds to $k$ iterations of the quantum gate.
Thus, the quantum circuit lengthened with each increase in the number of steps.
This problem is undesirable in NISQ, where noise is present.
The solution leverages that the unitary matrix $U$ can be diagonalized by the unitary matrix $V$.
\begin{align}
U^k(\bm{\theta}) =V(\bm{\phi})\Sigma(\bm{\gamma})^k V^{\dagger}(\bm{\phi})  
                 =V(\bm{\phi})\Sigma(k\bm{\gamma}) V^{\dagger}(\bm{\phi}) 
\end{align}
where $\Sigma$ is a diagonal matrix, and $\bm{\gamma}, \bm{\phi}$ are the parameters.
In addition, we added auxiliary qubits $\ket{\psi_{ancilla}}$ to the initial quantum state $\ket{\psi_0}$:
\begin{align}
\ket{\psi{(\bm{\theta})}}  = U^k(\bm{\theta})(\ket{\psi_0} \otimes \ket{\psi_{ancilla}}).
\end{align}
This operation created a correlation between the target and environmental systems.
The measurements were performed only on the target system.
These processes were similar to those of open quantum systems and improved the expressive power of quantum machine learning models.
The addition of ancillary qubits is sometimes described as a hidden layer in classical machine learning.

\subsection*{Numerical experiment}

\subsubsection*{Setting}

We applied the time-series quantum generative model to financial time-series data.
In particular, it generated data for forecasting and missing value completion.
We evaluated the potential utility of the time-series quantum generative model based on numerical experiments.

\begin{figure}[htbp]
\centering
\includegraphics[width=\linewidth]{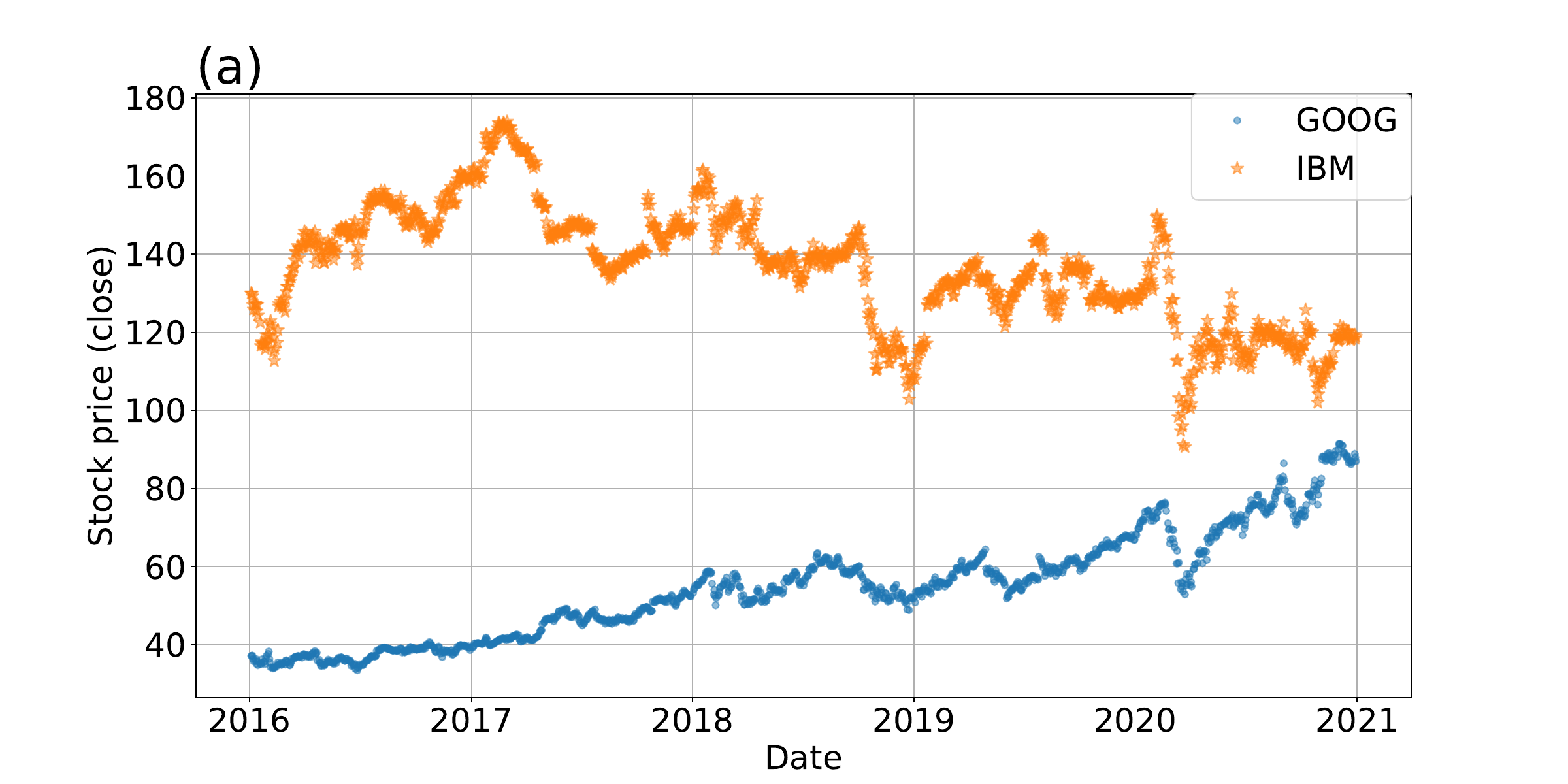}
\includegraphics[width=\linewidth]{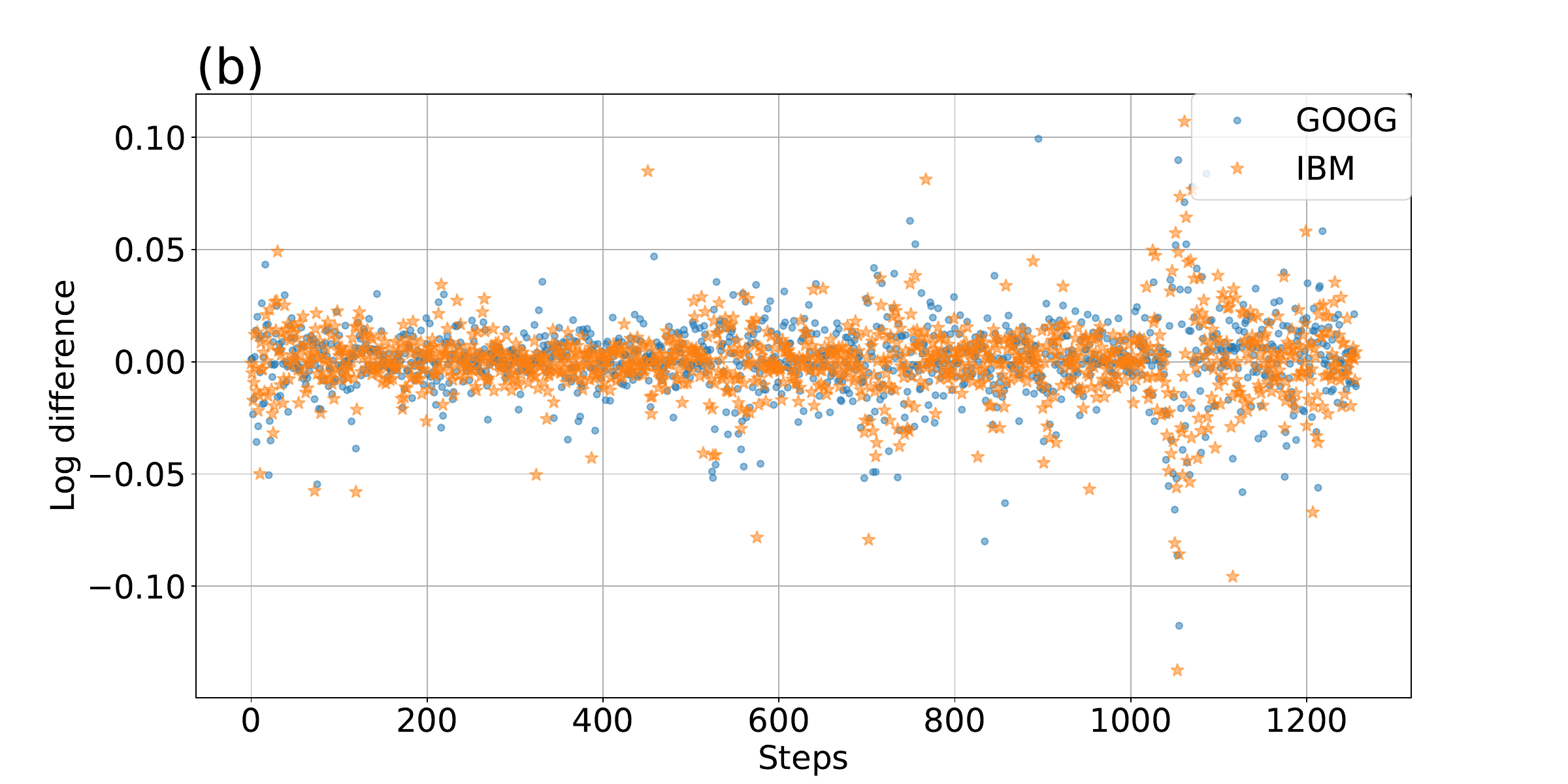}
\caption{(a) GOOG and IBM stock prices for 2016–2020 ( Close ). GOOG exhibits an increasing trend, whereas IBM exhibits a decreasing trend. (b) Logarithmic difference of (a. The logarithmic difference makes it a stationary time series.}
\label{SP}
\end{figure}

We used GOOGLE and IBM stock closing data for 2016--2020.
Figure \ref{SP} shows the stock prices of GOOGLE and IBM.
These data were obtained from Yahoo Finance\cite{YF}.
Because these time series are nonstationary, they were transformed into stationary time series for future forecasts.
For this purpose, we used the logarithmic difference $r_t$ defined as
\begin{align}
r_t  = \log{x_t} - \log{x_{t-1}}. \label{logdiff}
\end{align}
The conversion of stock prices into logarithmic differences yields a stationary time series.
By contrast, missing value completion does not use logarithmic differences, remains nonstationary, and uses data.
This is because many classical methods are unsuitable for nonstationary data. Moreover, this study was focused on verifying the performance of time-series generative models on such data.

GOOGLE and IBM stock prices were divided into five datasets by year.
The dataset for the year 20yy was denoted as D20yy.
In the case of future prediction, D20yy was used for training to predict the 10 steps not included in the dataset.
The prediction accuracy was compared using classical methods like vector autoregression (VAR) and long short-term memory (LSTM).
The number of lags in VAR was set to 50 or fewer, according to AIC.
LSTM has four layers and uses Adam with a learning rate 0.01 for parameter optimization. 
The accuracy of these predictions was evaluated in terms of MSE by performing the inverse transformation of Eq. (\ref{logdiff}).

\begin{figure}[htbp]
\centering
\includegraphics[width=60mm, angle=-90]{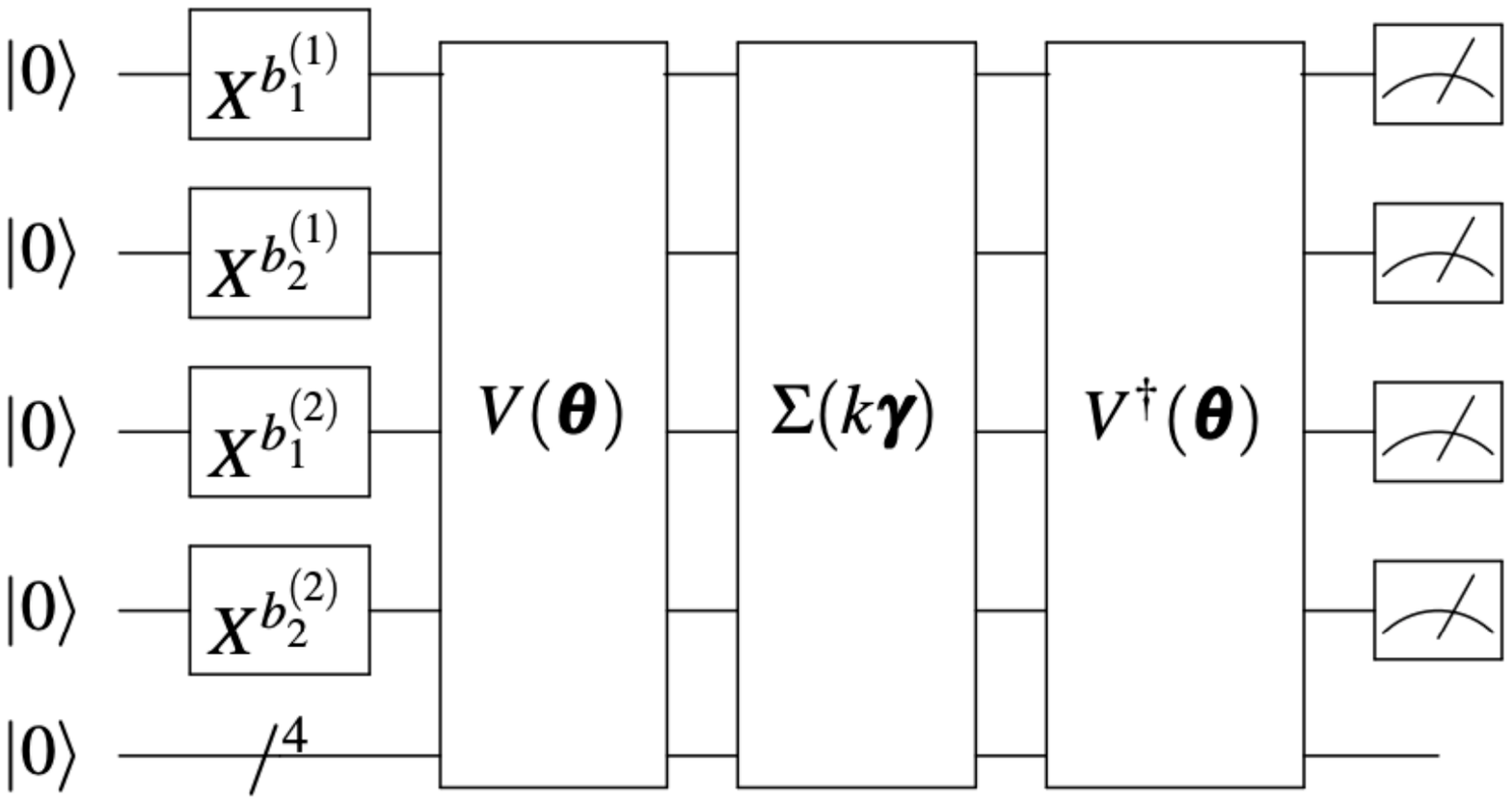}
\caption{Quantum circuits for numerical experiments. The discretized states of GOOG and IBM are encoded as bit strings as $\bm{b}_{GOOG}=b^{(1)}_1b^{(1)}_2$ and $\bm{b}_{IBM}=b^{(2)}_1b^{(2)}_2$, respectively.
Considering the number of $k$ steps, the state distribution after $k$ steps can be obtained through measurement.}
\label{QC}
\includegraphics[width=70mm, angle=-90]{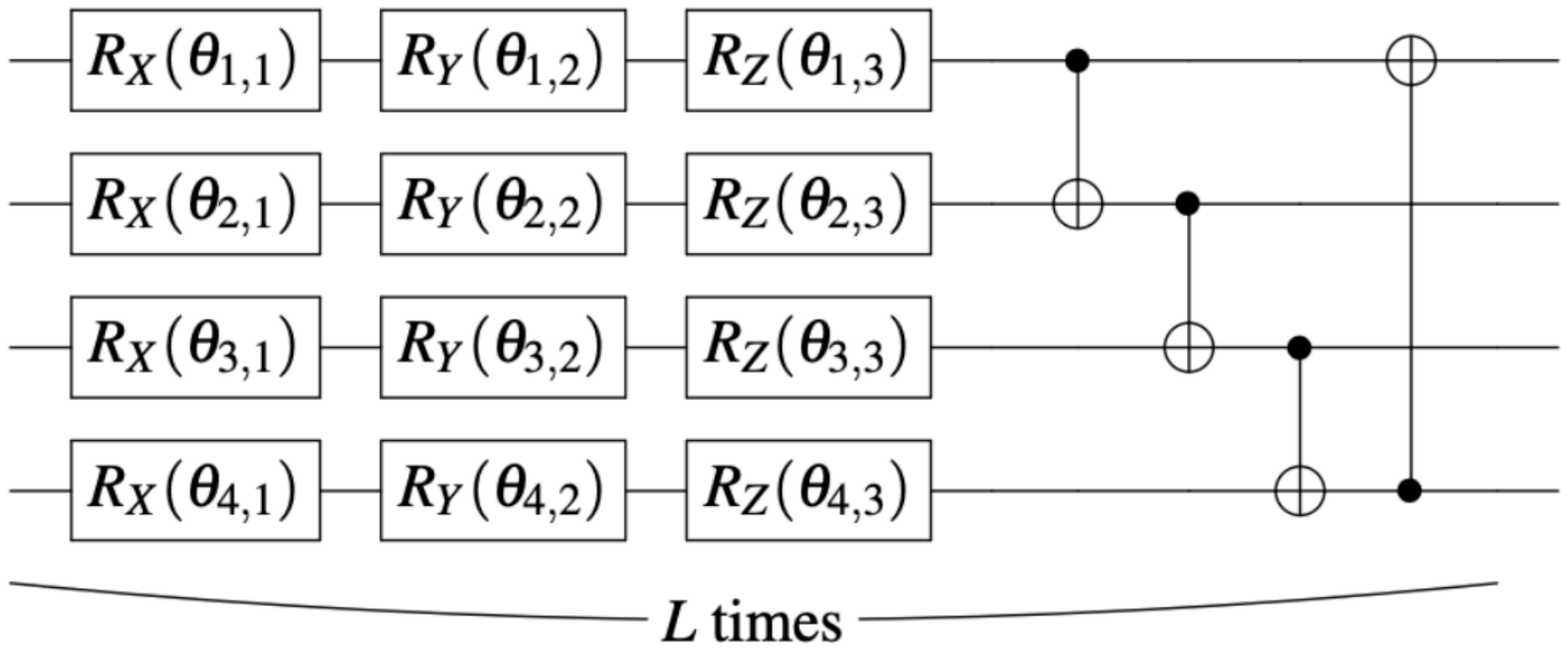}
\caption{Quantum circuit diagram of $V(\bm{\theta}$) for four qubits. This quantum circuit is called StronglyEntanglingLayers in PennyLane.
It comprises a single qubit rotation and an entanglement layer comprised of CNOT.}
\label{Vgate}
\includegraphics[width=55mm, angle=-90]{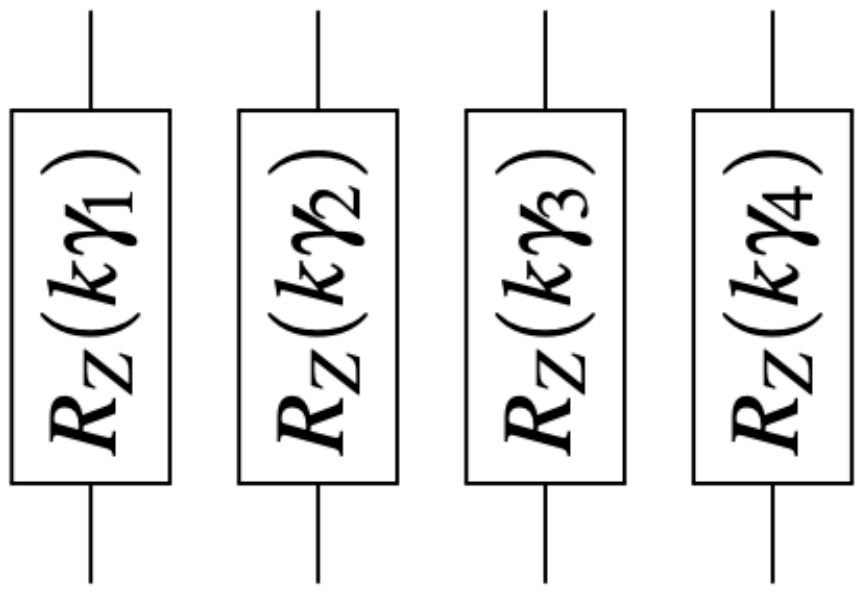}
\caption{Quantum circuit diagram of $\Sigma(\bm{\gamma}$) for four qubits. It comprises only Z rotation. This unitary operator is a diagonal matrix.}
\label{Sigmagate}
\end{figure}

In missing value completion, the data from the 50th--59th steps of D20yy were missing, and these data were completed using the quantum time-series generative model.
The data must be discretized for training using a time-series quantum generative model.
For this purpose, 25\% of the data in D20yy was assigned a discrete value $s=\{0, 1, 2, 3\}$ for each interval and discretized to $2^2$.
The quantum time-series generation model was executed on a simulator.
We used a standard PennyLane simulator\cite{PennyLane}.
The initial state $\ket{\psi_0}$ was encoded using X gates.
StronglyEntanglingLayers in PennyLane was used as a parameterized unitary operator $U(\bm{\theta})$ \cite{Ansatz}.
The parameterized operator $U(\bm{\theta})$ comprises a single qubit rotation and an entanglement layer.
The parameterized unitary operator can increase the number of parameters while improving expressiveness.
Here, four qubits were used as auxiliary qubits; thus, the quantum circuit had eight qubits.
We used Adam with a learning rate of 0.1 for parameter optimization.
Figure \ref{QC} shows the quantum circuit used in the numerical experiments, and Figure \ref{Vgate} shows the quantum circuit in StronglyEntanglingLayers.
Figure \ref{Sigmagate} shows the quantum circuit that formed the diagonal matrix $\Sigma$ used in the numerical experiments.

\subsubsection*{Results}

\begin{figure}[htbp]
\centering
\includegraphics[width=150mm]{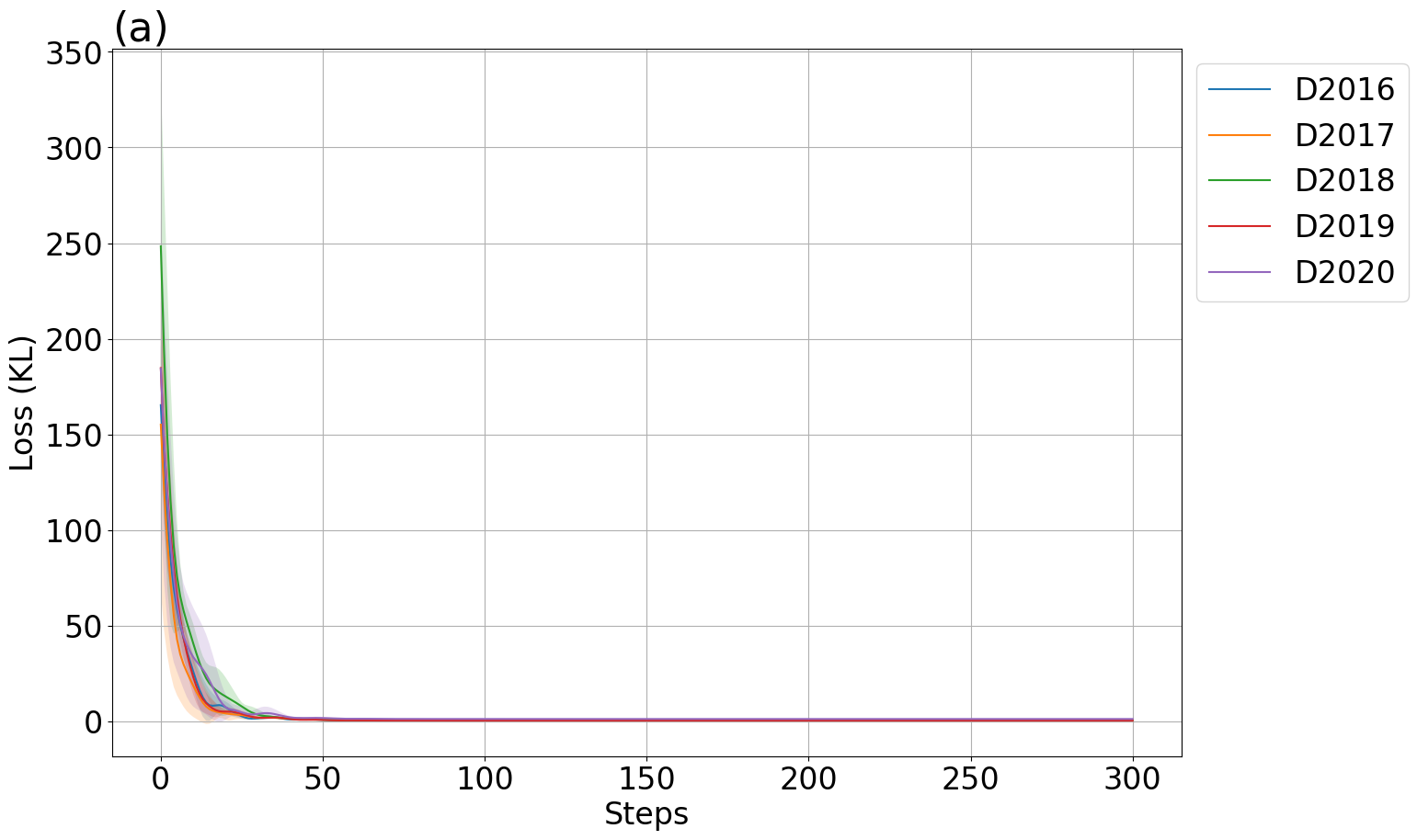}
\includegraphics[width=150mm]{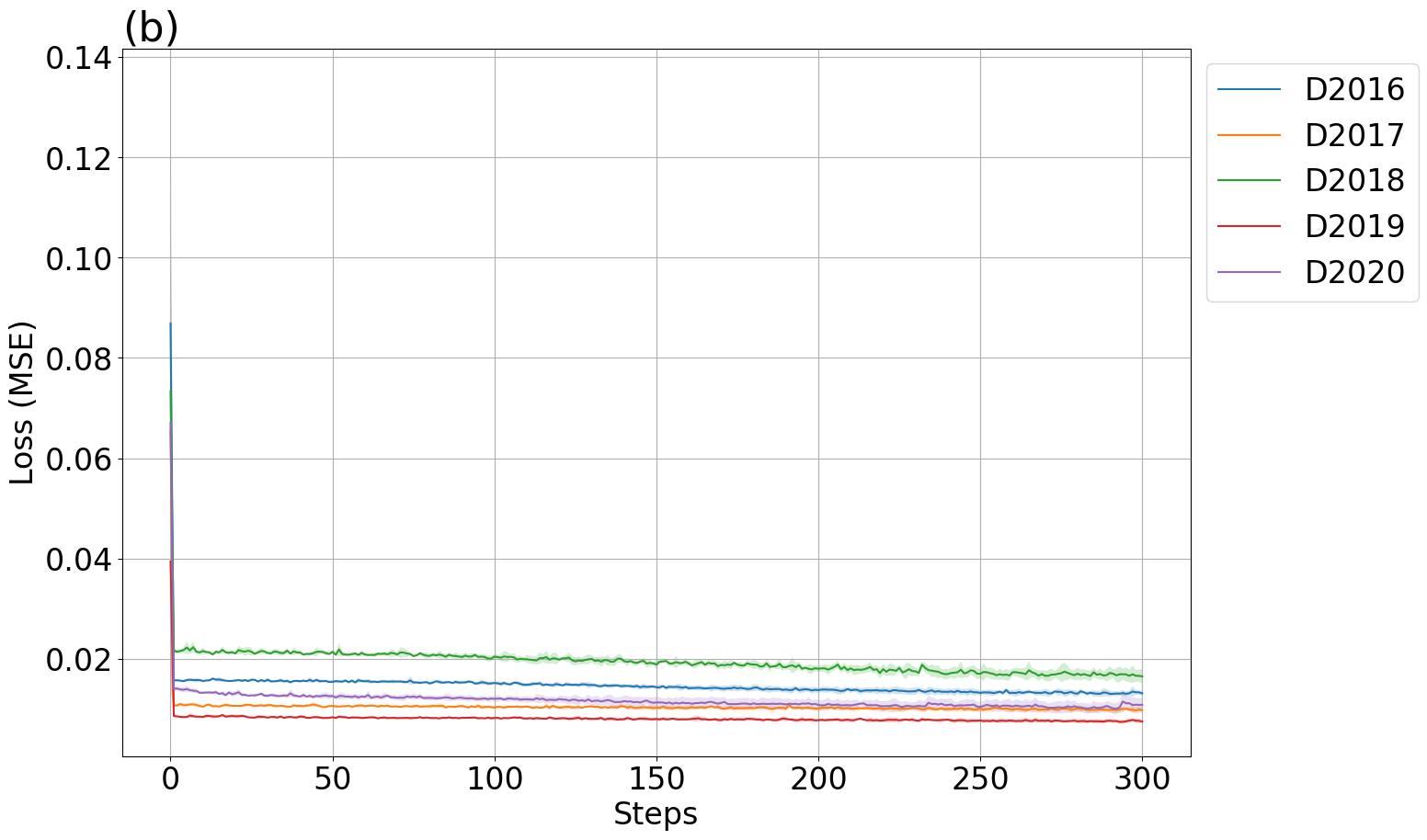}
\caption{Variation of loss at 300 steps. (a) and (b) show the time-series quantum generative model and the loss of LSTM. The time-series quantum generative model converges faster than LSTM.}
\label{Loss_pred}
\end{figure}

First, the results for future data generation are discussed.
Figure \ref{Loss_pred} shows the training loss for LSTM and the time-series quantum generative model.
We confirmed early convergence in both the models and datasets.
The number of learning steps required was the same for both models.
However, the learning rate of Adam was different.
Table \ref{pred} lists the results in terms of MSE for the values predicted by the time-series quantum generative model, LSTM, and VAR over ten steps.
The inverse transformation of the discretization restored the predictive data generated by the time-series quantum generative model.
Therefore, the reconstructed values were dependent on the training data statistics.
The logarithmic difference predicted by each model was transformed into time-series data using an inverse transformation.
\begin{table}[htbp]
  \begin{minipage}[t]{.5\textwidth}
    \begin{center}
      \begin{tabular}{lccc}
      (a)&  &  & \\
    \hline
    Dataset& t-QGEN & LSTM & VAR \\
    \hline \hline
D2016&2.57(0.467)&1.973(1.659)&$\bm{1.644}$\\
D2017&7.106(1.885)&$\bm{6.228}$(4.735)&8.159\\
D2018&3.266(1.078)&3.089(1.963)&$\bm{1.572}$\\
D2019&13.478(1.741)&16.083(12.636)&$\bm{10.756}$\\
D2020&3.223(35.884)&1.721(2.018)&$\bm{1.597}$\\
    \hline
  
  \end{tabular}
  \end{center}
  \end{minipage}
  \hfill
  \begin{minipage}[t]{.5\textwidth}
    \begin{center}
     \begin{tabular}{lccc}
     (b)&  &  & \\
    \hline
    Dataset& t-QGEN & LSTM & VAR \\
    \hline \hline
D2016&$\bm{2.591}$(32.508)&3.239(2.571)&3.594\\
D2017&86.932(38.354)&70.307(65.664)&$\bm{68.738}$\\
D2018&$\bm{36.636}$(3.345)&39.247(23.577)&38.045\\
D2019&10.65(4.379)&10.628(6.457)&$\bm{2.09}$\\
D2020&13.747(22.037)&$\bm{7.937}$(3.93)&14.403\\
    \hline
  \end{tabular}
    \end{center}
  \end{minipage}
  \caption{Mean (standard deviation) of future data generation results.
(a) and (b) show the predictions of GOOGLE and IBM, respectively.
t-QGE is a time-series quantum generative model.
The bold values denote the best result in the corresponding dataset.
In many cases, VAR provided the best outcome.}
\label{pred}
\end{table}

\begin{figure}[htbp]
\centering
\includegraphics[width=150mm]{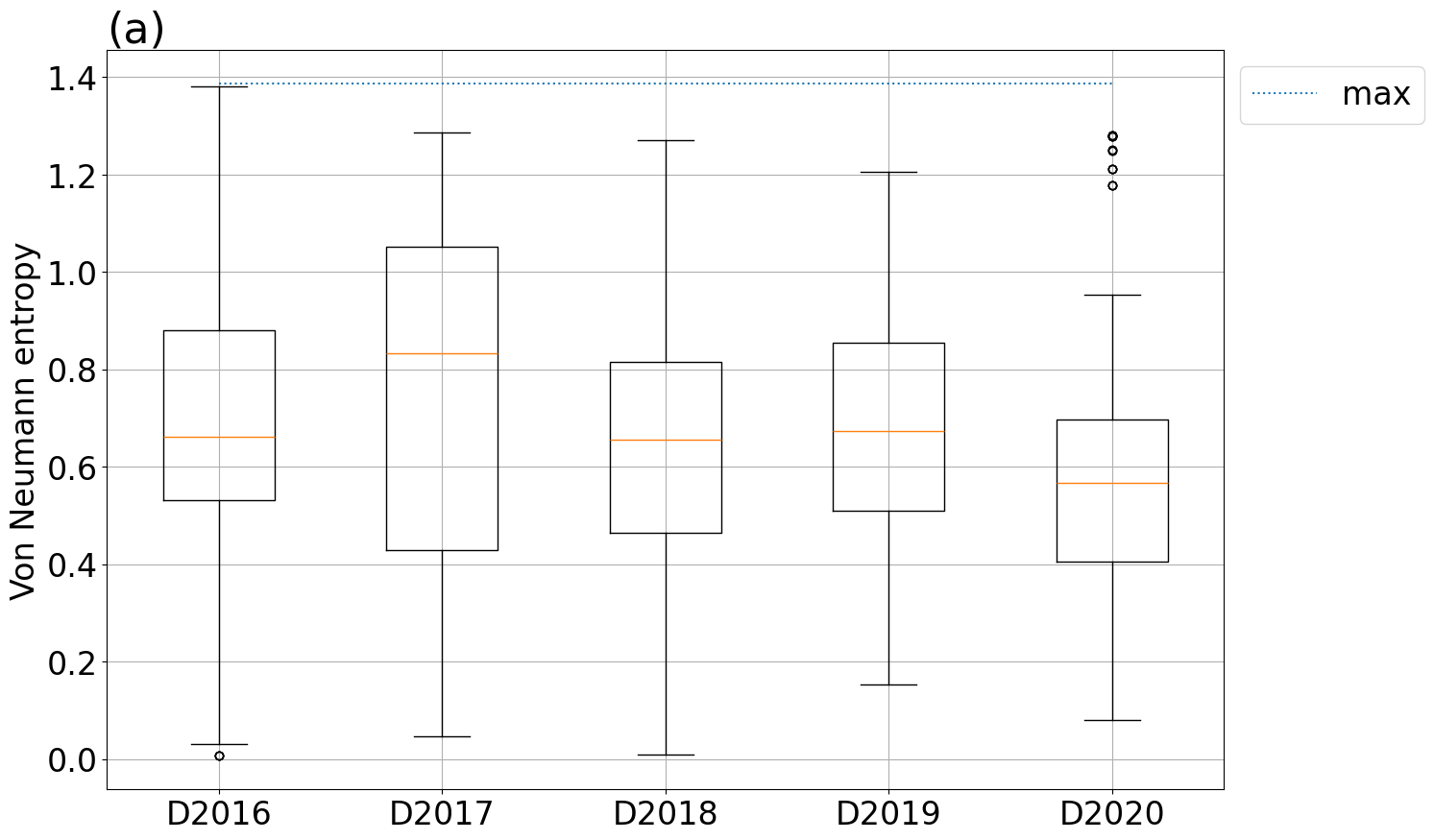}
\caption{von Neumann entropy between GOOGLE and the other systems.  Statistics were obtained by sampling from all states at $t=1–5$. Here max is the case of maximum entanglement. We confirmed the existence of entanglement in all datasets.}
\label{VNE}
\end{figure}

In most cases, VAR exhibited the best prediction accuracy because the logarithmic differencing results were as close as possible to the stationary time series, and the lag numbers were appropriately tuned.
Such results have been observed to appear in specific time-series forecasting datasets using the classical generative model\cite{QGEN}.
The time-series generative model and LSTM can modify the number of hidden layers, improving accuracy.
However, considering the overall results, we can assume that the quantum generative model can generate data with the same level of accuracy as that of LSTM.
Thus, the quantum time-series generation model requires fewer parameters than the classical methods.
The results generated by the quantum time-series generation model were discretized. 
As an evaluation function, we used the Manhattan distance $D_1$, which is defined as follows:
\begin{align}
D_1  = \frac{1}{10}\sum^{10}_{i=1} \abs{x_i - y_i}
\end{align}
where $x_i, y_i \in \{0,1,2,3\}$ are the true and predicted values of the $i$th step, respectively.
Table \ref{Manh_pred}, Figure \ref{Acc_pred}, and Table \ref{Acc_pred} present the Manhattan distance for each dataset and the increase in the Manhattan distance at each step.
\begin{table}[htbp]
  \centering
  \begin{tabular}{lccc}
    \hline
    Dataset& GOOG & IBM  \\
    \hline \hline
D2016&14.6 (4.72)&14.8 (1.72)\\
D2017&16.8 (2.79)&11.0 (1.41)\\
D2018&10.6 (1.85)&18.2 (2.4)\\
D2019&15.2 (2.32)&11.4 (3.01)\\
D2020&13.8 (2.32)&13.8 (2.32)\\
    \hline
  \end{tabular}
  \caption{Mean (standard deviation) of prediction accuracy of time-series quantum generative models by Manhattan distance $D_1$. 
As the Manhattan distance is smaller than 15 in most cases, significant prediction failures are considered to be rare.}
\label{Manh_pred}
\end{table}
\begin{figure}[htbp]
\centering
\includegraphics[width=150mm]{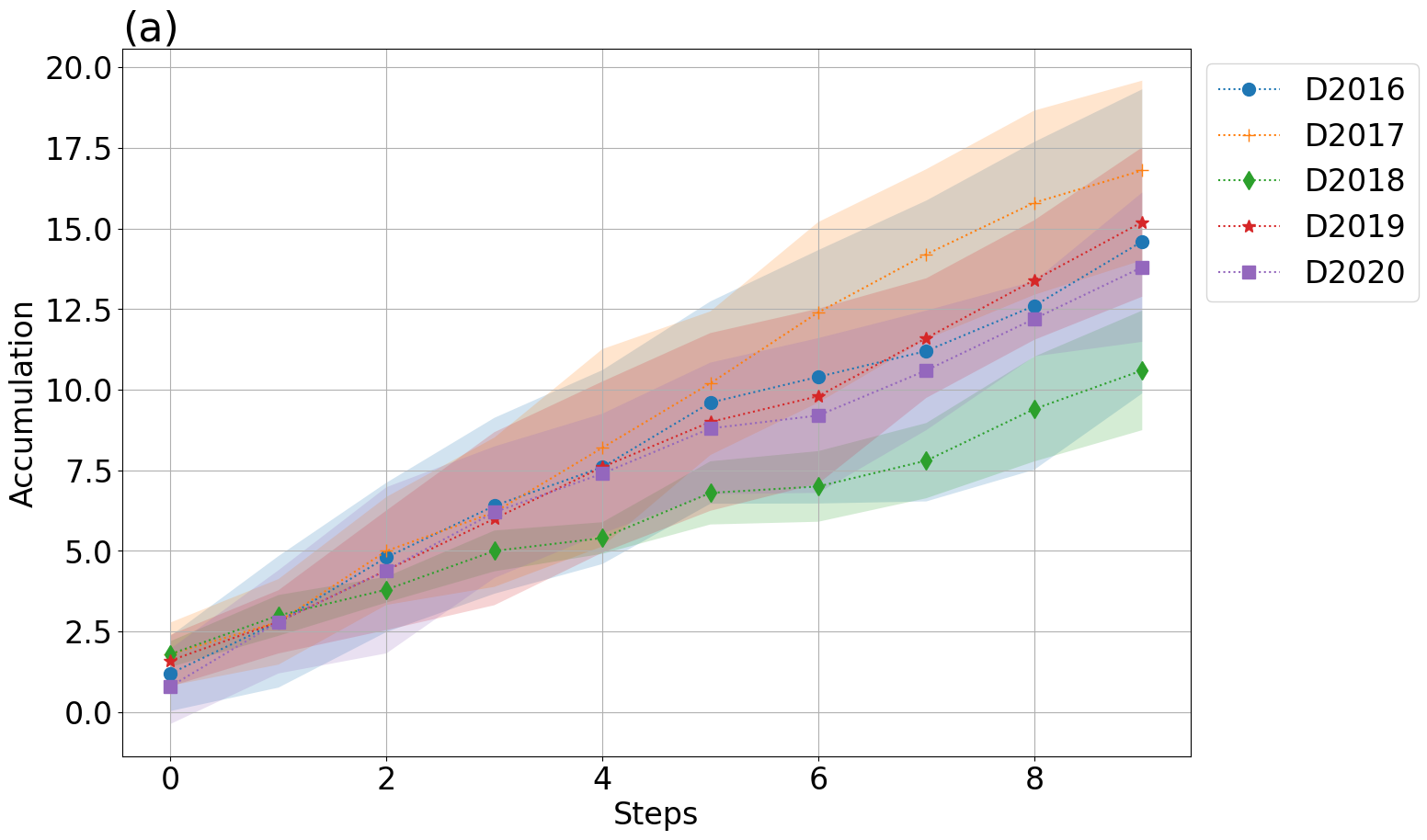}
\includegraphics[width=150mm]{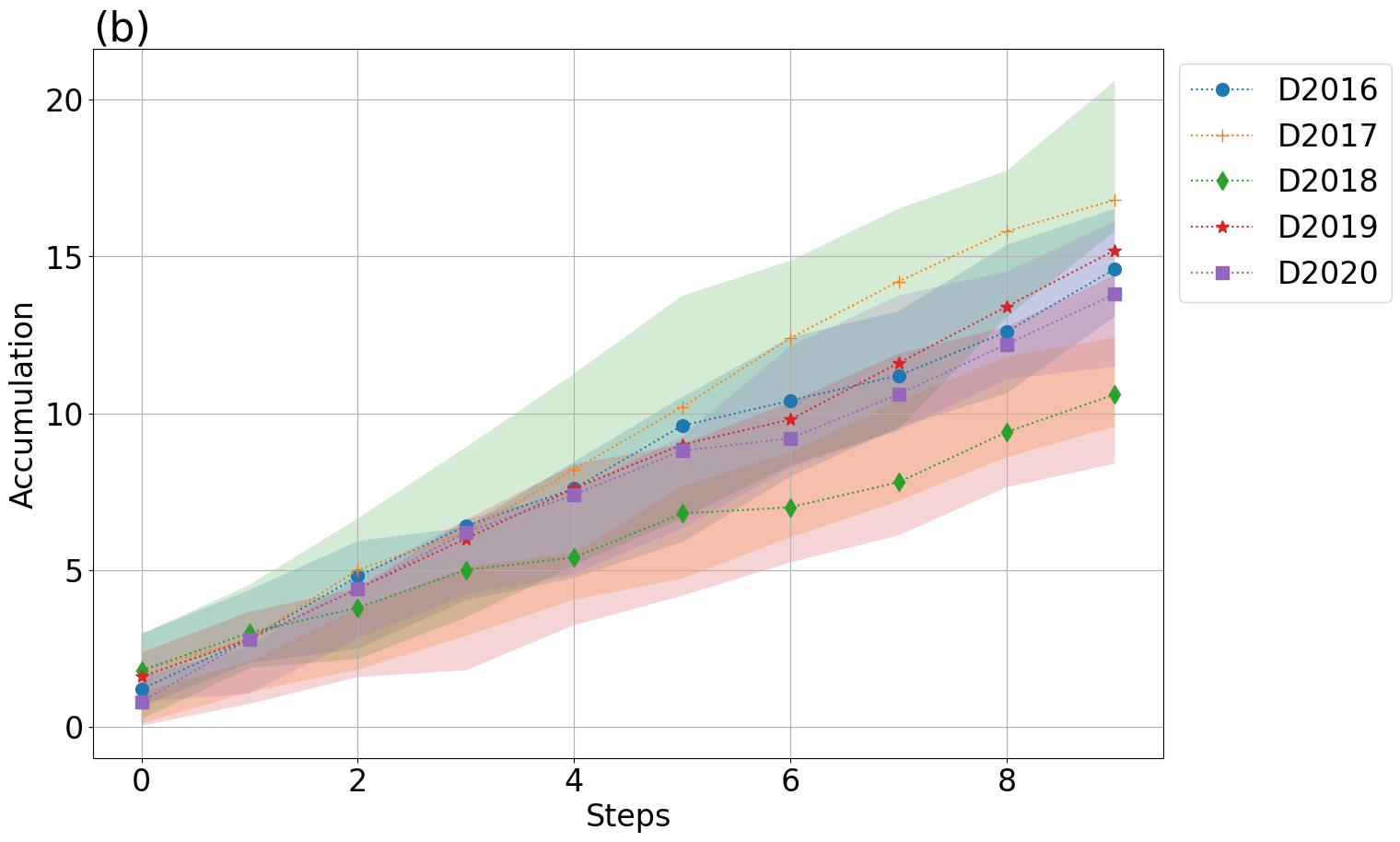}
\caption{Accumulation of Manhattan distance at each step. (a) and (b) show the results for GOOGLE and IBM, respectively. Both of these are linearly increasing.}
\label{Acc_pred2}
\end{figure}
\begin{table}[htbp]
  \centering
  \begin{tabular}{lccc}
    \hline
    Dataset& GOOG & IBM  \\
    \hline \hline
D2016&1.43 (0.99)&1.44 (0.993)\\
D2017&1.77 (0.995)&1.18 (0.997)\\
D2018&0.92 (0.987)&1.79 (0.993)\\
D2019&1.49 (0.997)&1.15 (0.998)\\
D2020&1.36 (0.988)&1.49 (0.992)\\
    \hline
  \end{tabular}
  \caption{Slope (coefficient of determination) for cumulative Manhattan distance. In all cases, it is a linear function. Therefore, the magnitude of the error generated at each step is approximately the same.}
 \label{Acc_pred} 
\end{table}
For most datasets, the Manhattan distance was less than 15.
The increase in the Manhattan distance was linear and unlikely to result in significant errors at each step.
These results support the idea that the quantum time-series generation model can capture the trends of the correlated time series.
The von Neumann entropy between GOOGLE and the other systems at $t=1$–$t=5$ is shown in Figure. \ref{VNE}.
The results suggest that entanglement aids in the capture of the correlation between GOOGLE and IBM.

\begin{figure}[htbp]
\centering
\includegraphics[width=150mm]{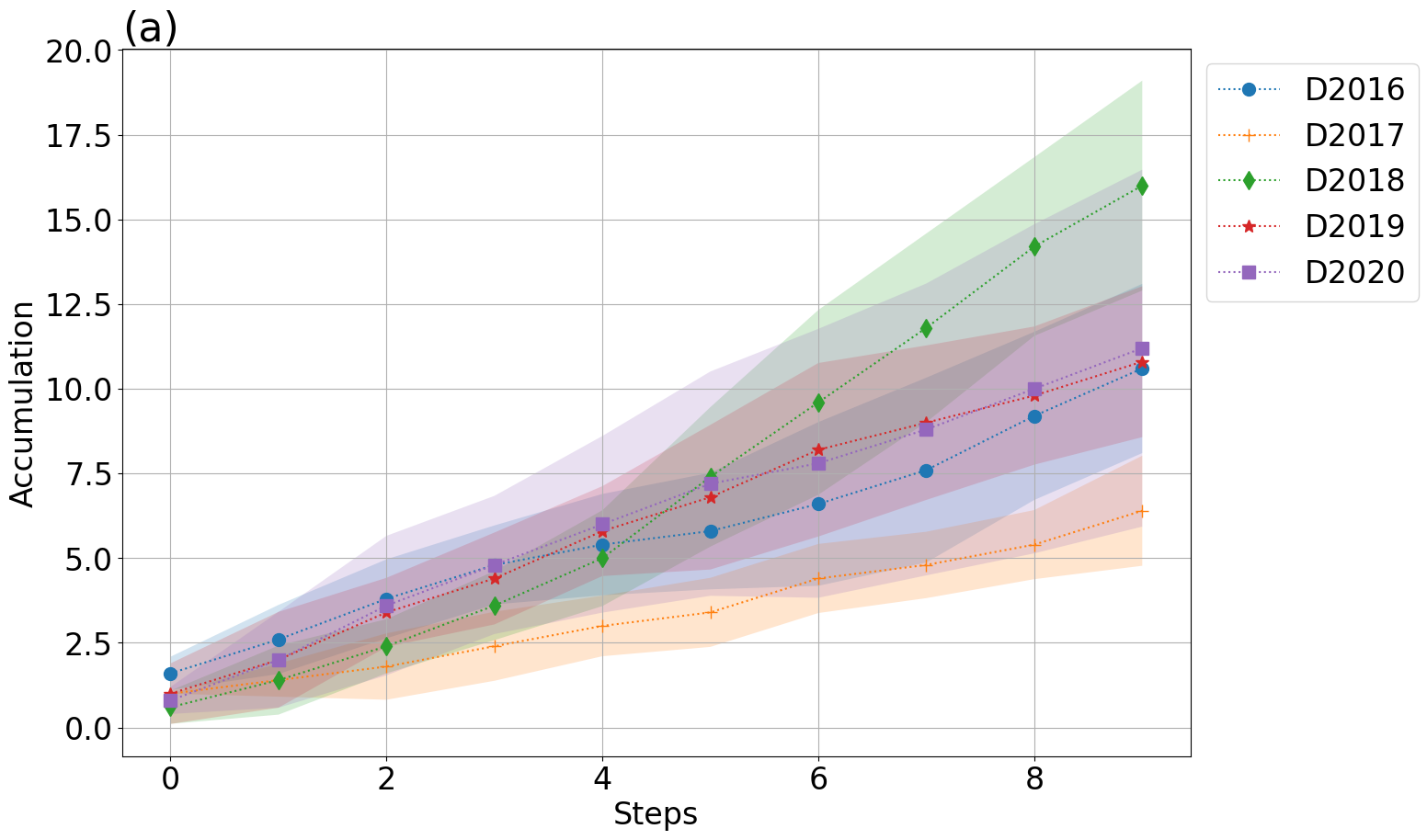}
\includegraphics[width=150mm]{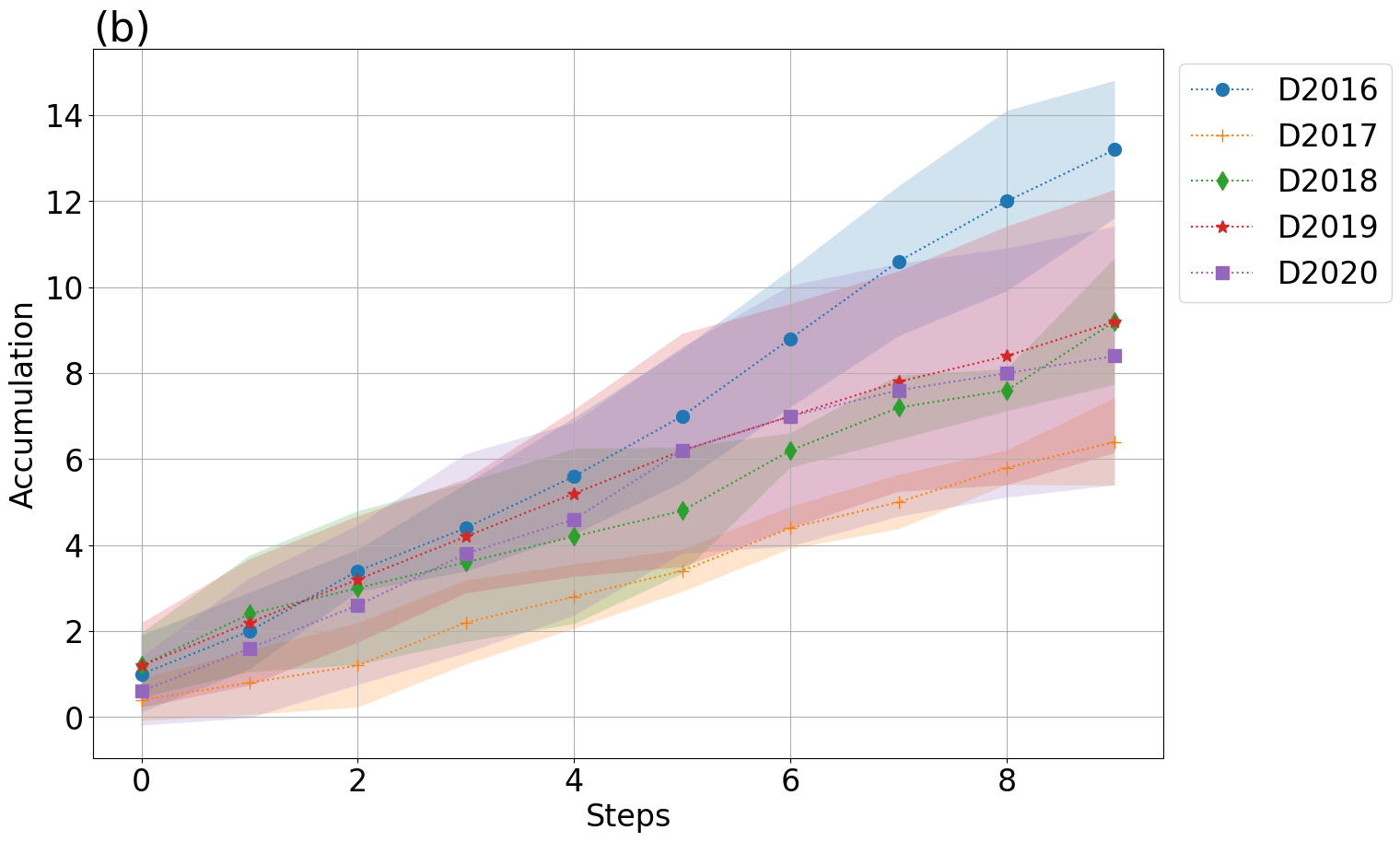}
\caption{Accumulation of Manhattan distance at each step. (a) and (b) show the results for GOOGLE ($L=1$) and GOOG($L=3$), respectively. Both of these are linearly increasing.}
\label{Acc_comp}
\end{figure}

Next, we discuss the results of completing the missing values.
Numerical experiments were performed on the time-series quantum generative model with $L=1$ and $L=3$.
We evaluated the results using discrete values because inverse transformation depends on the training data and the discretization method.
Table \ref{comp}, Figure \ref{Acc_comp}, and Table \ref{Manh_comp} present the accuracy of completion based on the Manhattan distance and the increase in Manhattan distance at each step.

\begin{table}[htbp]
  \begin{minipage}[t]{.5\textwidth}
    \begin{center}
     \begin{tabular}{lccc}
     (a)&&\\
    \hline
    Dataset& GOOG & IBM  \\
    \hline \hline
D2016&10.6 (2.5)&9.8 (1.47)\\
D2017&6.4 (1.62)&15.0 (5.22)\\
D2018&16.0 (3.1)&9.0 (3.29)\\
D2019&10.8 (2.23)&7.4 (2.42)\\
D2020&11.2 (5.27)&14.4 (4.32)\\
    \hline
  \end{tabular}
  \end{center}
  \end{minipage}
  \hfill
  \begin{minipage}[t]{.5\textwidth}
    \begin{center}
     \begin{tabular}{lccc}
     (b)&&\\
    \hline
    Dataset& GOOG & IBM  \\
    \hline \hline
D2016&13.2 (1.6)&11.4 (2.58)\\
D2017&6.4 (1.02)&15.6 (3.38)\\
D2018&9.2 (1.47)&12.2 (3.19)\\
D2019&9.2 (3.06)&10.0 (1.41)\\
D2020&8.4 (3.01)&14.0 (2.83)\\
    \hline
  \end{tabular}
    \end{center}
  \end{minipage}
  \caption{Mean (standard deviation) of Completion of missing values results.
(a) and (b) show the completion results when the layers of the quantum circuit are $L=1$ and $L=3$, respectively.
In both cases, the Manhattan distance is approximately 10.
In addition, the standard deviation becomes smaller as the number of layers increases.}
\label{comp}
\end{table}

\begin{table}[htbp]
  \begin{minipage}[t]{.5\textwidth}
    \begin{center}
     \begin{tabular}{lccc}
    (a)&&\\
    \hline
    Dataset& GOOG & IBM  \\
    \hline \hline
D2016&0.92 (0.979)&1.3 (0.986)\\
D2017&0.59 (0.989)&1.67 (0.982)\\
D2018&0.9 (0.994)&0.98 (0.979)\\
D2019&1.11 (0.995)&0.23 (0.975)\\
D2020&1.13 (0.993)&1.41 (0.986)\\
    \hline
  \end{tabular}
  \end{center}
  \end{minipage}
  \hfill
  \begin{minipage}[t]{.5\textwidth}
    \begin{center}
     \begin{tabular}{lccc}
    (b)&&\\
    \hline
    Dataset& GOOG & IBM  \\
    \hline \hline
D2016&1.4 (0.993)&1.25 (0.988)\\
D2017&0.7 (0.994)&1.65 (0.999)\\
D2018&0.84 (0.984)&0.98 (0.979)\\
D2019&0.9 (0.994)&0.98 (0.979)\\
D2020&0.92 (0.975)&1.37 (0.996)\\
    \hline
  \end{tabular}
    \end{center}
  \end{minipage}
  \caption{Slope (coefficient of determination) of the cumulative Manhattan distance for each layer of the quantum circuit for $L=1$ and $L=3$. For all results, a linear function is observed.}
\label{Manh_comp}
\end{table}

We considered appropriate complementation successful because the Manhattan distance is less than 10 for both layers, $L = 1$ and $L=3$.
In addition, as in future prediction, the increase in error at each step was linear.
As the number of layers increased, the standard deviation decreased, and the extent to which accuracy scattered from one training to another decreased.
This result indicates that even with one layer, completion often succeeded, but the certainty increased with the number of layers.
Therefore, more layers are preferred to obtain reliable results.

Because the data used here were nonstationary, it was challenging to use VAR or LSTM.
Therefore, this may be a valuable case for time-series quantum generative models.

\section*{Conclusion}

This study adopted a time-series quantum generative model for financial data.
For future data predictions, we compared them with typical classical methods: LSTM and VAR.
Consequently, we confirmed that the time-series quantum generative model yielded the same accuracy as LSTM with fewer parameters.
By evaluating the Manhattan distance, we confirmed that the time-series quantum generative model could capture the trend of the correlated time series.
These results may be attributed to the functioning of entanglement.
The completion of missing values using the Manhattan distance confirmed the effectiveness of the time-series quantum generative model.
In addition, we obtained results suggesting that the number of layers in the quantum circuit could be controlled by learning to control scattering.

We confirmed that the time-series quantum generative model applied to stationary and nonstationary data through numerical experiments.
In particular, because nonstationary data are not well-suited to typical classical methods, the ability to generate data with a small number of parameters may be advantageous.
However, the coarseness of the discretization must be fine to compute true predictions from the inverse transform of the predicted discrete values.
Moreover, this operation should be performed cautiously because it increases the number of quantum bits and computational complexity.
The proposal of an appropriate discretization and restoration method to solve such problems is challenging.
It would also be interesting to consider the modification of loss functions.
In recent years, loss functions based on optimal transport have been successfully applied in quantum generative models \cite{Opt}.
By using such loss functions, it may be possible to reduce the training time without losing accuracy by fully using mini-batches.

\section*{Acknowledgements}

This study was supported by JSPS KAKENHI Grant No. 23H01432.
This study was financially supported by the Public\verb|\|Private R\&D Investment Strategic Expansion PrograM (PRISM) and programs for bridging the gap between R\&D and IDeal society (Society 5.0) and Generating Economic and social value (BRIDGE) from the Cabinet Office.

\section*{Author contributions statement}

S. O. conceived and conducted the experiments and analyzed the results.  
M. A. and M. O. supervised the study. 
All the authors reviewed the draft manuscript and critically revised it for intellectual content. 
All authors have approved the final version of the manuscript for publication.

\end{document}